\documentclass[acmsmall,screen,nonacm]{acmart}
\AtBeginDocument{%
  }

\setcopyright{none}
\usepackage{algorithm}
\usepackage{algorithmic}
\usepackage{tcolorbox}
\tcbuselibrary{skins,breakable}
\renewcommand\footnotetextcopyrightpermission[1]{}
\setcopyright{none}
\author{Mohammad Wathiq Soualhi}
\affiliation{%
  \institution{Independent Researcher}
  \city{Kuala Lumpur}
  \country{Malaysia}
}
\email{wathiqclaw@gmail.com}

\author{Basem Suleiman}
\affiliation{%
  \institution{University of New South Wales}
  \city{Sydney}
  \country{Australia}
}
\email{b.suleiman@unsw.edu.au}
\authorsaddresses{}
\begin{document}
%% THis was added for arxiv
\pagestyle{plain}
\title[Co-Modifying Application Code and Deployment Infrastructure]{Co-Modifying Application Code and Deployment Infrastructure: A Causal Study of AI Agent Pull Requests}

\begin{abstract}
Current software engineering agents are predominantly evaluated inside isolated sandboxes against synchronous, local tests. However, deploying features in collaborative repositories requires modifying continuous integration pipelines, container configurations, and provisioning manifests alongside application logic. Because agents lack deployment awareness, attempting these cross-domain modifications introduces unverified dependencies into asynchronous integration pipelines. To isolate the causal integration overhead of dual-scope pull requests (PRs), which modify both application code and deployment infrastructure, we evaluated 26,999 PRs across two large-scale datasets (AIDev v5 and MOSAIC-3M) encompassing six agent architectures and an aligned human baseline. By contributing a deterministic four-tier file taxonomy and an exact multi-dimensional strata-matching framework, we isolated integration metrics under conditional exchangeability while controlling for application volume. Currently, agents operate primarily as application coders or maintenance workers, with pure maintenance tasks capturing 17.70\% to 25.41\% of agent-authored PRs. Dual-scope modifications represent only 4.48\% to 4.55\% of agentic PRs, yet they trigger bimodal integration outcomes. In the first mode, explicitly reviewed contributions experience protracted delays, adding a median penalty of +13.13 hours and multiplying time-to-merge by $3.24\times$, driven mechanically by continuous integration and provisioning manifests. In the second mode, maintainer gatekeeping exhibits a structural bypass: even for external contributors lacking repository write access, agent identity reduces the odds of human review by 63.7\% compared to human peers. Consequently, 43.1\% to 56.2\% of dual-scope PRs bypass inspection and merge silently in a median of under 31 minutes. These uninspected modifications compromise the build verification oracle by creating a circular verification failure, establishing the need for deployment-aware benchmarks and strict path-based access controls.
\end{abstract}






\maketitle

% ------------------------------------------------------------------------------
% MODULAR PAPER SECTIONS
% ------------------------------------------------------------------------------
\section{Introduction}\label{sec:introduction}

AI coding agents increasingly author and submit source code contributions across collaborative software repositories~\cite{aidev_v5, mosaic3m_2026}. Current evaluation benchmarks assess these agents primarily within isolated containerized sandboxes, spanning micro-level code synthesis (HumanEval~\cite{chen2021evaluating}), repository-scale issue resolution (SWE-bench~\cite{swebench2024}, DevBench~\cite{li2024devbench}), and specialized infrastructure automation (DevOps-Gym~\cite{tang2026devopsgym}). While these frameworks effectively measure offline code synthesis and local test resolution, they isolate code generation from asynchronous continuous integration pipelines and shared repository governance. In production environments, however, operationalizing new application logic frequently requires concurrently modifying continuous integration (CI/CD) pipelines, container configurations, and declarative infrastructure-as-code (IaC) manifests~\cite{jiang2015coevolution, gallaba2018use, sharma2016code}. In this work, we conceptualize these infrastructure configurations not as disjointed coding tasks, but as the \textit{operational context} required to successfully deploy functional application code into production.

Current agent architectures operate with a localized synthesis bias: they generate code without proactive mechanisms to anticipate downstream deployment constraints, asynchronous integration pipelines, or human maintainer scrutiny. When an agent attempts to concurrently modify application code and operational infrastructure within the same contribution---a pattern we formalize as a \textit{dual-scope pull request} (PR)---it transmits unverified cross-domain dependencies into the live repository. This dynamic not only replicates historical pre-DevOps organizational silos~\cite{bass2015devops}, but also introduces socio-technical vulnerabilities. By automating infrastructure modifications without established governance protocols, agents risk creating a \textit{productivity illusion}: rapid code generation is either offset by multi-day delays in the maintainer review queue, or it bypasses human inspection entirely, establishing uninspected supply-chain vectors.

Despite the growing prevalence of agent-authored code contributions, existing empirical software engineering literature lacks a rigorous evaluation of this cross-domain integration overhead. Prior large-scale mining studies evaluate agent performance predominantly through coarse binary merge acceptance rates, raw code churn proxies, or subjective natural-language task categorizations~\cite{yoshioka2026, peralta2026, ehsani2026}. Furthermore, by analyzing observational repository data without enforcing conditional exchangeability, existing studies cannot disentangle the true integration penalty from structural confounders---such as functional application size, architectural archetype, and test verification rigor. Consequently, the software engineering community lacks causal evidence regarding three critical dimensions of agent contributions: (1) the operational scope of cross-domain infrastructure modifications, (2) the mechanical latency bottlenecks these modifications induce, and (3) the extent to which human oversight bifurcates between agent-authored and human-authored dual-scope contributions.

To address this empirical gap, we formulate three research questions:
\begin{itemize}\setlength{\itemsep}{1pt}\setlength{\parskip}{0pt}\setlength{\parsep}{0pt}
    \item[\textbf{RQ1:}] What is the operational scope of agent-authored PRs, how do they co-modify application code alongside deployment infrastructure, and which subdomains do they target?
    \item[\textbf{RQ2:}] How does the merge latency of agent-authored dual-scope PRs compare to that of equivalent application-only PRs, and which infrastructure subdomains drive integration bottlenecks?
    \item[\textbf{RQ3:}] How does the inclusion of deployment infrastructure alter human maintainer oversight of agent-authored PRs compared to human developer baselines, and does repository governance moderate this intervention?
\end{itemize}

To investigate these questions, we construct a harmonized cross-validation cohort spanning 26,999 PRs across two large-scale datasets: AIDev v5 ($N = 21,855$)~\cite{aidev_v5} and MOSAIC-3M ($N = 5,144$, comprising 2,570 agent PRs and 2,574 human PRs)~\cite{mosaic3m_2026}. Across the two cohorts, we evaluate six agent architectures alongside an aligned human baseline, restricting analysis to mature repositories with over 100 stars. To eliminate subjective classification noise and upstream natural language processing (NLP) attrition, we develop a deterministic, four-tier sequential file-domain taxonomy that categorizes modified files based on path semantics and active authoring status. Furthermore, to isolate the causal integration penalty of infrastructure co-modifications, we implement a greedy 1:1 nearest-neighbor matching design that pairs dual-scope and application-only PRs within exact strata (controlling for agent architecture, architectural archetype, test maturity, and language ecosystem) while matching closely on application file volume.

Our empirical investigation reveals that while dual-scope infrastructure modifications represent a narrow operational slice (4.48\% to 4.55\% of agent-authored PRs) compared to non-functional repository maintenance chores (17.70\% to 25.41\%), they introduce statistically significant integration friction. Specifically, co-modifying infrastructure escalates median merge latency by +13.13 hours in AIDev v5, multiplying total time-to-merge by $3.24\times$. This delay is driven primarily by continuous integration manifests and declarative system provisioning, which multiply integration times by $3.8\times$ and $2.7\times$, respectively.

Furthermore, maintainer oversight exhibits a structural bifurcation based on author identity. While human review rates remain stable when developers co-modify infrastructure (72.2\% to 73.4\%), agent-authored dual-scope PRs trigger volatile oversight---increasing external review rates from 47.5\% to 53.1\% in AIDev v5, while dropping them from 52.5\% to 36.2\% in MOSAIC-3M. Even among external contributors with zero repository write access, agent identity reduces the odds of human review by 63.7\%. Consequently, repository workflows default to a bimodal integration trap: 43.1\% to 56.2\% of agent-authored dual-scope PRs bypass external review entirely to merge silently in a median of under 31 minutes, whereas explicitly reviewed contributions stall in the integration queue for 15.59 to 49.80 hours.

This paper makes the following scholarly contributions:
\begin{itemize}\setlength{\itemsep}{1pt}\setlength{\parskip}{0pt}\setlength{\parsep}{0pt}
    \item \textbf{Operational Characterization of Agent Contributions:} The first dual-dataset empirical evaluation quantifying dual-scope PRs and repository maintenance chores across 26,999 PRs, six agent architectures, and an aligned human baseline.
    \item \textbf{Deterministic Framework:} An extensible four-tier file taxonomy and exact 4D strata matching algorithm that controls for functional application file volume and eliminates reliance on subjective, high-attrition natural language task labels.
    \item \textbf{Causal Identification of Integration Friction:} Quantitative isolation of the +13.13-hour / $3.24\times$ merge latency penalty, decomposing the specific continuous integration and provisioning manifests that mechanically drive integration delays.
    \item \textbf{Socio-Technical Governance \& Supply-Chain Security:} Empirical evidence demonstrating that agent-authored PRs face a 63.7\% human review deficit independently of repository administrative permissions, establishing a quantifiable supply-chain vulnerability.
\end{itemize}

\section{Empirical Study Design}\label{sec:methodology}

We evaluate the integration overhead of co-modifying deployment infrastructure alongside application code using a multidimensional observational study design across two large-scale datasets, establishing conditional exchangeability through temporal bounding, deterministic file taxonomy, and exact strata matching.

\subsection{Target Datasets and Repository Quality Filtering}\label{sec:datasets}

We construct our empirical cohorts using two large-scale datasets of agentic PRs and repository metadata: AIDev v5 and MOSAIC-3M. Table~\ref{tab:datasets} summarizes the dataset provenance, experimental controls, and the progressive cohort filtering funnel applied to establish our evaluated populations.

For AIDev v5, we use the \texttt{AIDev-pop} subset. The \texttt{AIDev-pop} subset inherently enforces a repository popularity threshold of strictly greater than 100 stars. To address known perils of mining uncurated GitHub repositories that are predominantly personal or inactive~\cite{kalliamvakou2014promises}, we enforce this threshold across both cohorts to isolate mature, collaboratively maintained codebases. The dataset profiles six discrete agent architectures: OpenAI Codex~\cite{chen2021evaluating}, Claude Code~\cite{anthropic2025claudecode}, GitHub Copilot~\cite{ziegler2022copilot}, Devin~\cite{cognition2024devin}, Cursor~\cite{anysphere2024cursor}, and Google Jules~\cite{google2025jules}.

MOSAIC-3M profiles five of these exact architectures (Codex, Claude, Copilot, Devin, and Jules, yielding 2,570 cleaned agent PRs) alongside an aligned human baseline cohort ($N = 2,574$, comprising the 5,144 evaluated PRs in Table~\ref{tab:datasets}). This architectural parity ensures that our cross-dataset validations in RQ1 and RQ2 evaluate identical agents. Conversely, dataset divergences establish necessary boundaries for specific evaluations: AIDev v5 uniquely captures Cursor, while the MOSAIC-3M human baseline serves as the mandatory reference baseline for the cross-sectional interaction analysis in RQ3. To enforce cross-dataset structural parity and align with the \texttt{AIDev-pop} baseline, we apply a symmetrical repository quality filter to MOSAIC-3M, restricting analysis strictly to base repositories possessing greater than 100 stars.
\begin{table}
  \caption{Cross-Dataset Structural Parity and Cohort Funnel Matrix}
  \label{tab:datasets}
  \Description{Table mapping dataset provenance, structural boundaries, experimental controls, and the progressive cohort filtering funnel from raw PRs down to the cleaned evaluated populations across the AIDev-pop subset and MOSAIC-3M.}
  \footnotesize
  \begin{tabular}{p{3.8cm} p{3.4cm} p{3.4cm}}
    \toprule
    \textbf{Dimension / Pipeline Stage} & \textbf{AIDev v5 (\texttt{pop})} & \textbf{MOSAIC-3M} \\
    \midrule
    \multicolumn{3}{l}{\textit{\textbf{A. Dataset Scope \& Repository-Level Curation}}} \\
    Native Collection Timeline & Jan 2025 -- Mar 2026 & Jun 2025 -- Aug 2025 \\
    Profiled Agent Architectures & {\raggedright Codex, Claude, Copilot, Cursor, Devin, Jules\par} & {\raggedright Codex, Claude, Copilot, Devin, Jules\par} \\
    Human Baseline & None & Included \\
    Repository Popularity Threshold & $> 100\star$ (Inherent to \texttt{pop}) & $> 100\star$ (Applied filter) \\
    Curated Repositories ($>100\star$) & 25,402 & 4,553 (of 50,515) \\
    \midrule
    \multicolumn{3}{l}{\textit{\textbf{B. PR Cohort Funnel (Created Jun--Jul 2025, Merged $\le$ Aug 2025)}}} \\
    Raw Corpus PRs & 361,296 & 111,969 \\
    Target Window PRs & 26,870 & 5,206 \\
    Upstream API File Omissions & 5,015 (18.66\%) & 62 (1.19\%) \\
    \textbf{Cleaned Evaluated PRs ($N$)} & \textbf{21,855} & \textbf{5,144} \\
    \bottomrule
  \end{tabular}
\end{table}
\subsection{Temporal Cohort Controls and Data Attrition Audit}\label{sec:temporal}

We isolate a strict temporal creation cohort consisting of PRs authored between June 1, 2025, and July 31, 2025. We enforce a 31-day observation buffer on this cohort, mandating that all evaluated PRs possess a terminal merge timestamp on or before August 31, 2025. This boundary alignment eliminates the left-censoring of stale, legacy PRs authored by unrefined early-2025 agent models. Simultaneously, the 31-day observation buffer mitigates right-censoring, commonly known as fast-merge bias, ensuring that complex dual-scope PRs are not disproportionately truncated from the dataset simply because they require extended review cycles. Empirical latency distributions confirm that 99.41\% of AIDev v5 PRs and 98.27\% of MOSAIC-3M PRs successfully merge within 30 days. This validates that our 31-day calendar observation buffer (the entirety of August) captures the full terminal merge distribution while eliminating follow-up truncation.

Before advancing to feature extraction, we account for known upstream data attrition caused by GitHub API payload limits. As explicitly noted in the AIDev dataset documentation, standard extraction pipelines often omit patch data and file-level changes for exceptionally large PRs because the GitHub API refuses to provide them. This API ceiling causes downstream data collection to yield incomplete or entirely missing file arrays. An empirical audit of AIDev v5 reveals that this omission escalates continuously across the 15-month timeline: rising from 0.00\% in January 2025, surpassing 30\% by October 2025, and peaking at 65.64\% by March 2026, totaling 117,292 permanently omitted PRs. This accelerating data loss renders longitudinal analysis unreliable and dictates that temporal bounding is a methodological necessity. Consequently, we abandon a full 15-month evaluation in favor of the strictly aligned June-to-August cross-validation cohort.

Within our isolated target window, 18.66\% of merged AIDev v5 PRs (5,015 out of 26,870) lack usable file paths in the associated commit tables. This omission is systematically concentrated in a single agent track: 99.78\% of the permanently dropped records (5,004 PRs) belong to Claude Code, representing an unrecoverable 94.50\% data loss for that specific agent model. Attempts to reconstruct the missing file paths via regex parsing of the raw Git commit \texttt{patch} strings yielded zero recoverable records, confirming that the upstream API size limits caused the payload to omit the patch data entirely alongside the file arrays.

% \begin{figure}
%   \includegraphics[width=0.75\linewidth]{samples/figures/placeholder_figure.png}
%   \caption{Temporal Attrition Audit: Escalating upstream extraction data loss in AiDev v5}
%   \label{fig:attrition}
%   \Description{A line chart showing data loss percentage starting at 0 percent in January 2025 and rising steadily to over 65 percent by March 2026.}
% \end{figure}

To prevent this upstream truncation from distorting empirical prevalence rates and causal matching, we implement an explicit payload safeguard for the MOSAIC-3M dataset and use it as our cross-validation baseline. We strictly drop 56 MOSAIC-3M PRs where the length of the extracted file array is less than the GitHub API \texttt{changed\_files} integer, guaranteeing payload integrity. Consequently, the MOSAIC-3M dataset exhibits 99.88\% file-level completeness across the aligned temporal cohort, dropping exactly 6 out of the remaining 5,150 PRs. This structural parity guarantees that our subsequent file classification and causal matching pipelines operate on observable repository data.
\subsection{Deterministic File Domain Taxonomy}\label{sec:taxonomy}

To evaluate the integration overhead of infrastructure co-modifications, prior empirical methodologies rely on NLP to classify PR task types. For instance, early iterations of the AIDev dataset deployed an LLM-based pipeline to auto-classify PRs into Conventional Commit categories~\cite{aidev_v5}. Conversely, datasets like MOSAIC-3M provide no native classification pipeline, requiring researchers to heuristically group raw PR labels, titles, and linked issue tags~\cite{mosaic3m_2026}. An empirical audit of the AIDev v5 \texttt{pr\_task\_type} table reveals a chronological sparsity anomaly. Because this NLP pipeline was discontinued during the v5 timeline refresh, valid classification coverage declines from 66.32\% in July 2025 to just 1.13\% by August 2025. Attempting to harmonize this temporally truncated metadata with MOSAIC-3M's unstandardized textual fields introduces systemic missing-data bias.

We discard subjective NLP task classifications in favor of a deterministic file-level taxonomy, achieving 100\% classification coverage and rigorous cross-dataset parity. We extract all modified file paths, change statuses, and change volumes across both datasets. For AIDev v5, we apply chronological commit deduplication; while we retain the final commit state of each modified file to prevent intermediate revisions from inflating file counts, we mathematically aggregate the gross line churn across all intermediate commits to preserve the total rework volume. For MOSAIC-3M, we achieve parity by parsing and deduplicating the nested file arrays via unique path sets. Furthermore, we evaluate the file change status to enforce action integrity, distinguishing actively authored infrastructure from pure file deletions. To ground our classification rules in observable developer behavior, we first executed an exploratory extraction of over 368,000 deduplicated file paths across both datasets, profiling the top 50 most frequent extensions, exact filenames, and root directories. Informed by these empirical distributions, we subsequently process every deduplicated file through a deterministic, mutually exclusive four-tier sequential classifier. Once a file satisfies a tier's conditions, it is immediately classified, preventing overlapping definitions:

\begin{enumerate}
    \item \textbf{Tier 1: Unambiguous Infrastructure Manifests.} The classifier first isolates unambiguous infrastructure manifests regardless of their repository directory, mapping them into three core subcategories: \texttt{CI/CD} pipelines (e.g., \texttt{.github/workflows/}), \texttt{Containers} (e.g., \texttt{Dockerfile}), and declarative IaC Provisioning templates (e.g., \texttt{.tf}, \texttt{cdk.json}).
    \item \textbf{Tier 2: Test Isolation.} The classifier subsequently intercepts testing assets via explicit test directories (e.g., \texttt{tests/}, \texttt{spec/}) and prefix/suffix filename patterns. Executing this step before Tier 3 prevents test automation scripts and mock fixtures from artificially inflating infrastructure or application metrics.
    \item \textbf{Tier 3: Context-Dependent Infrastructure.} The classifier assigns the \texttt{Automation} subcategory to procedural scripts (e.g., \texttt{.sh} or \texttt{.ps1} scripts located in \texttt{scripts/} or \texttt{tools/}) and maps directory-based IaC (e.g., \texttt{terraform/} directories) to the \texttt{Provisioning} subcategory.
    \item \textbf{Tier 4: Structural Application and Configuration.} Remaining files are evaluated for functional characteristics. Core application source code (e.g., \texttt{.py}, \texttt{.go}) is tagged independently from application configurations (e.g., \texttt{package.json}, \texttt{pyproject.toml} within \texttt{src/} or \texttt{app/}) to distinguish core application logic from application configuration adjustments. To prevent false positives, standard source code located in utility directories (e.g., \texttt{scripts/}) and generic configurations outside of recognized application directories are intercepted and reclassified as auxiliary maintenance, alongside pure documentation and lockfiles.
\end{enumerate}

Following file-level classification, we aggregate the file-level data to establish the PR scope. A PR is categorized as a dual-scope PR if it modifies a minimum of one valid application file, one valid infrastructure file, and contains a total of at least two modified files. PRs modifying exclusively application code or exclusively infrastructure code are designated as single-domain baselines (application-only and infrastructure-only PRs, respectively). PRs containing zero application code and zero infrastructure code (e.g., pure documentation updates or lockfile bumps) are classified as auxiliary maintenance. These maintenance PRs are quantified during our prevalence evaluation (RQ1) but are subsequently excluded from the causal matching pipelines (RQ2 and RQ3), which strictly evaluate the integration of functional development.

\subsection{Causal Identification and Exact Strata Matching}\label{sec:matching}

To isolate the integration penalty of infrastructure co-modifications, we mitigate observed selection bias by enforcing strict conditional exchangeability. We cannot naively compare dual-scope PRs to single-domain PRs, as dual-scope PRs inherently modify a larger total volume of files, introducing size-based confounding into merge latency and review evaluations. We implement a Greedy 1:1 Nearest-Neighbor Matching design without replacement, formalized in Algorithm~\ref{alg:matching}, pairing each dual-scope treatment unit with an application-only control unit across four exact strata dimensions:

\begin{itemize}
    \item \textbf{Agent Architecture:} We enforce direct, within-agent pairing (e.g., matching a Copilot-authored treatment PR strictly to a Copilot-authored control PR). This controls for confounding driven by intrinsic baseline speed and behavioral differentials among the LLMs.
    \item \textbf{Architectural Archetype:} We match on primary structural location across eight repository archetypes: Core/Library, Frontend/App, Backend/API, Root (Top-Level), Cross-Domain/Full-Stack, Examples/Samples, Vendor/Dependencies, and Root/Other. To prevent the treatment condition itself from shifting the archetype classification, we calculate the primary repository location exclusively using the modified application files within the PR. If multiple directories tie for the highest file count, the PR is deterministically assigned a Cross-Domain/Full-Stack archetype to ensure multi-domain modifications are matched symmetrically.
    \item \textbf{Test Verification Maturity:} We hold verification overhead constant by matching strictly on test file inclusion. PRs are paired conditionally based on whether they explicitly introduce or modify test automation assets (identified via our Tier 2 classifier), ensuring equivalent structural verification scope.
    \item \textbf{Language Ecosystem:} We restrict matches to identical technology stacks across nine partitioned ecosystems (\texttt{Node/Frontend}, \texttt{Python}, \texttt{Compiled Systems}, \texttt{JVM}, \texttt{.NET}, \texttt{Interpreted Backend}, \texttt{Mobile/Native UI}, \texttt{Ops/Scripting}, and \texttt{Other/Niche}) derived from the repository's primary language metadata, eliminating build-system and runtime variance.
\end{itemize}

Within these multidimensional strata, we calculate the standardized Euclidean distance between candidates based exclusively on functional development breadth. We define this metric as the $z$-score of the log-transformed application file volume. We sort the treatment units descending by application volume to prioritize matching larger PRs before the control pool is depleted. The matching algorithm pairs each treatment PR to the nearest available control PR, enforcing a maximum distance caliper of $\epsilon \le 0.28$ standard deviations. Aligned with established causal matching literature recommending calipers within the $0.20\text{--}0.25\sigma$ neighborhood~\cite{austin2011psm, rosenbaum1985constructing}, this threshold accommodates the discrete integer distribution of file counts while retaining common support. Matched control units are immediately removed from the candidate pool to ensure 1:1 discrete pairing.

\begin{algorithm}
\footnotesize
\caption{Greedy 1:1 Nearest-Neighbor Caliper Matching within Exact Strata}
\label{alg:matching}
\begin{algorithmic}[1]
\REQUIRE Treatment pool $T$, Control pool $C$, Strata dimensions $S$, Matching covariate $D$, Caliper $\epsilon$
\ENSURE Set of matched pairs $M$
\STATE $M \leftarrow \emptyset$
\STATE $\mathcal{B} \leftarrow \text{Partition}(T \cup C, S)$ \COMMENT{Group into exact strata blocks}
\FOR{\textbf{each} stratum block $B \in \mathcal{B}$}
    \STATE $T_B \leftarrow \{x \in B \mid x \in T\}$
    \STATE $C_B \leftarrow \{x \in B \mid x \in C\}$
    \STATE Sort $T_B$ descending by covariate $D$ \COMMENT{Prioritize high-volume PRs}
    \FOR{\textbf{each} $t \in T_B$}
        \STATE $C_{candidates} \leftarrow \{c \in C_B \mid |t.D - c.D| \le \epsilon\}$
        \IF{$C_{candidates} \neq \emptyset$}
            \STATE $c^* \leftarrow \arg\min_{c \in C_{candidates}} |t.D - c.D|$ \COMMENT{Ties broken deterministically}
            \STATE $M \leftarrow M \cup \{(t, c^*)\}$
            \STATE $C_B \leftarrow C_B \setminus \{c^*\}$ \COMMENT{Match without replacement}
        \ENDIF
    \ENDFOR
\ENDFOR
\RETURN $M$
\end{algorithmic}
\end{algorithm}

Prior to matching execution, we validated the common support (positivity) assumption by auditing 4D strata cell occupancy, confirming sufficient control pool density to pair treatment units while strictly enforcing the $\epsilon \le 0.28$ standard deviation caliper. Enforcing this caliper pruned 29.82\% (292 of 979) of dual-scope PRs in AIDev v5 and 33.88\% (41 of 121 agent PRs) in MOSAIC-3M due to extreme functional sizes exceeding common support, retaining 687 and 80 balanced matched pairs, respectively. The resulting matched pairs allow us to estimate the Average Treatment Effect on the Treated (ATT). For continuous velocity outcomes in RQ2, we evaluate paired differences using the Wilcoxon signed-rank test~\cite{wilcoxon1945individual} with Pratt's zero-method for ties~\cite{pratt1959remarks} and Cliff's delta effect sizes~\cite{cliff1993dominance}, complemented by a log-normal Accelerated Failure Time (AFT) survival model~\cite{kalbfleisch2002statistical}. To properly model heavily right-skewed duration distributions, this parametric model evaluates fully observed uncensored durations conditional on successful PR integration, while applying Huber--White cluster-robust sandwich standard errors~\cite{white1980heteroskedasticity} clustered on matched pairs. To evaluate binary human intervention outcomes for RQ3, we evaluate discordant transitions via the exact McNemar test~\cite{mcnemar1947note} and model review probability using both pooled and governance-stratified logistic regressions. These models adjust for functional application line churn and repository popularity, which are excluded from the matching caliper to prevent dimensional collapse while eliminating post-treatment bias.

To ensure construct validity across these statistical models, we formalize three dependent variables. First, we define \textit{Merge Latency} as the elapsed time in hours between PR creation and integration, floored at 0.001 hours (3.6 seconds) to prevent numerical divergence in our Log-Normal survival models for near-instantaneous automated merges. We explicitly discard commit-based rework metrics, as our empirical construct validity audit revealed that 68.62\% of AIDev v5 PRs and 42.00\% of MOSAIC-3M PRs undergo squash-merging into single-commit merges, systematically obscuring intermediate revision histories. Second, to evaluate \textit{Review Friction}, we extract PR discussion comments and formal review iterations. To prevent automated CI/CD status checks from artificially inflating review volume, we execute a deterministic regex sanitization engine to purge bot notifications and service accounts (eliminating 63.80\% of raw comments in AIDev v5 and 67.96\% in MOSAIC-3M). Finally, we isolate external maintainer oversight by filtering out author self-comments. Because a diagnostic audit confirmed that author self-comments do not systematically skew intervention metrics (shifting the review status of only 2.49\% of AIDev and 3.23\% of MOSAIC PRs), this isolation allows us to formally and safely define a \textit{Silent Integration} as a PR that merges with exactly zero external human comments and zero external review iterations.

Finally, to establish an empirical baseline for human review behavior in RQ3, we independently apply this exact matching protocol to the MOSAIC-3M human cohort. This human-to-human matching employs the identical application volume distance metric and remaining exact strata, omitting only the agent architecture dimension, retaining 79 balanced matched pairs (pruning 14.13\% of treatment units) to isolate the baseline integration friction of human dual-scope contributions.
\section{Empirical Results}\label{sec:results}

We present our empirical evaluation of agentic dual-scope contributions across three dimensions of the integration pipeline: operational prevalence (RQ1), causal integration latency (RQ2), and human gatekeeping bifurcation (RQ3).

\subsection{RQ1: Empirical Scale and the ``Chore Agent'' Phenomenon}\label{sec:rq1}

\begin{figure}
 \small \includegraphics[width=0.75\linewidth]{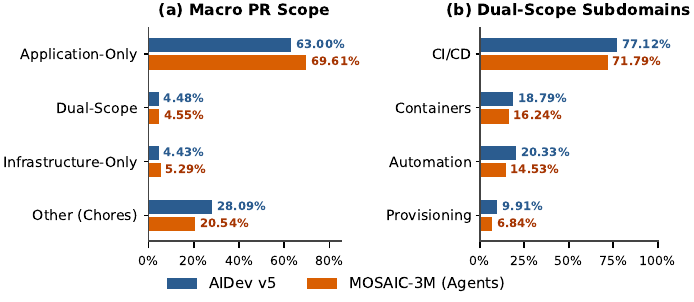}
\caption{Operational Landscape of Agentic Pull Requests: (a) Macro PR scope distribution across cleaned cohorts; (b) Infrastructure subdomains modified within dual-scope PRs (non-mutually exclusive).}
  \label{fig:rq1_landscape}
\Description{Two-panel visualization: Panel (a) is a grouped horizontal bar chart comparing AIDev v5 and MOSAIC-3M across macro PR scopes, showing application-only PRs at 63 to 70 percent, dual-scope PRs at roughly 4.5 percent, and maintenance chores at 21 to 28 percent. Panel (b) shows dual-scope subdomains, dominated by CI/CD manifests at 72 to 77 percent, followed by Containers and Automation (15 to 20 percent each), and Provisioning (7 to 10 percent).}
\end{figure}

We evaluate the operational scope of agent PRs and the prevalence of cross-domain infrastructure modifications across the cleaned cohorts of 21,855 AIDev v5 PRs and 5,144 MOSAIC-3M PRs (partitioned into 2,570 agentic PRs and 2,574 human PRs). As illustrated in Figure~\ref{fig:rq1_landscape}(a), the operational landscape across both cohorts is predominantly driven by isolated application feature development (application-only PRs), accounting for 63.00\% of AIDev v5 and 69.61\% of MOSAIC-3M PRs. In contrast, dual-scope PRs represent a narrow operational intersection, capturing 4.48\% (979) of AIDev v5 and 4.55\% (117) of agentic PRs in MOSAIC-3M. Isolated infrastructure-only PRs represent a comparable minor fraction at 4.43\% and 5.29\%, respectively.

These dual-scope PRs represent active engineering rather than auxiliary file deletion or trivial manifest bumps. Action integrity audits reveal that 97.45\% of AIDev v5 and 99.01\% of MOSAIC-3M dual-scope PRs involve the active authoring, addition, or modification of infrastructure files. Furthermore, 91.11\% (AIDev) and 86.63\% (MOSAIC) of these PRs exhibit strict dual-scope behavior, directly modifying core application source code alongside infrastructure, rather than executing trivial modifications restricted solely to application configuration files (e.g., \texttt{package.json}). To evaluate the intentionality of these modifications, we measure dual-scope intensity by the absolute volume of infrastructure files modified. Empirical observations reveal that agentic infrastructure co-modifications are highly targeted: in AIDev v5, the median infrastructure file count per dual-scope PR is 2.0, with 47.0\% modifying exactly one infrastructure file. In MOSAIC-3M, the median is 1.0, with 63.9\% modifying exactly one file. When agents modify exactly one infrastructure file, it is overwhelmingly a CI/CD manifest (74.1\% in AIDev v5 and 65.9\% in MOSAIC-3M), demonstrating discrete adjustments to core deployment pipelines rather than incidental directory changes.

Decomposing this dual-scope slice across operational domains (categories are non-mutually exclusive), Figure~\ref{fig:rq1_landscape}(b) demonstrates that infrastructure co-modifications are heavily skewed toward continuous integration and deployment (CI/CD) pipelines (77.12\% in AIDev v5 and 71.79\% in MOSAIC-3M). In contrast, procedural automation scripts (20.33\% and 14.53\%), containerization configurations (18.79\% and 16.24\%), and declarative system provisioning templates (9.91\% and 6.84\%) record significantly lower participation. Operationally, this disparity indicates that current agents are predominantly deployed to patch existing workflow pipelines rather than to architect or provision core infrastructure.

Table~\ref{tab:rq1_agent_prevalence} summarizes the complete agent-level dual-scope prevalence, demonstrating that Copilot and Devin exhibit the highest rates of infrastructure integration across both cohorts, whereas OpenAI Codex consistently records the lowest propensity for cross-domain modification. In the AIDev v5 cohort, Claude Code exhibits an apparent prevalence of 17.87\%. However, as established in Section~\ref{sec:temporal}, this is a mathematical artifact of survivorship bias caused by the upstream API omission of 5,004 Claude Code PRs (representing 94.50\% data loss for this architecture). Assuming an absolute minimum bound where all 5,004 omitted PRs were single-domain, the true prevalence drops to 0.98\%. Imputing the unbiased 6.09\% baseline observed in the MOSAIC-3M Claude cohort yields an expected AIDev prevalence of 6.74\%. Consequently, we rely exclusively on MOSAIC-3M for baseline empirical observations of the Claude architecture.
\begin{table}
  \caption{Cross-Dataset Agent-Level Dual-Scope Prevalence (PRs Created Jun--Jul 2025)}
  \label{tab:rq1_agent_prevalence}
  \Description{Table comparing dual-scope PR prevalence across agent architectures side-by-side between AIDev v5 and MOSAIC-3M cohorts, reporting total PRs, dual-scope counts, and percentage prevalence.}
  \footnotesize
  \begin{tabular}{l rrr rrr}
    \toprule
    & \multicolumn{3}{c}{\textbf{AIDev v5}} & \multicolumn{3}{c}{\textbf{MOSAIC-3M}} \\
    \cmidrule(lr){2-4} \cmidrule(lr){5-7}
\textbf{Author / Architecture} & \textbf{Total PRs} & \textbf{Dual-Scope} & \textbf{Rate (\%)} & \textbf{Total PRs} & \textbf{Dual-Scope} & \textbf{Rate (\%)} \\
    \midrule
    Claude Code & 291 & 52 & 17.87\%$^*$ & 903 & 55 & 6.09\% \\
    Copilot & 2,675 & 413 & 15.44\% & 505 & 23 & 4.55\% \\
    Devin & 751 & 91 & 12.12\% & 858 & 33 & 3.85\% \\
    Google Jules & 27 & 3 & 11.11\% & 26 & 1 & 3.85\% \\
    Cursor & 1,176 & 103 & 8.76\% & \multicolumn{3}{c}{\textit{Not Tracked}} \\
    OpenAI Codex & 16,935 & 317 & 1.87\% & 278 & 5 & 1.80\% \\
    \midrule
        \textbf{All Agents (Combined)} & \textbf{21,855} & \textbf{979} & \textbf{4.48\%} & \textbf{2,570} & \textbf{117} & \textbf{4.55\%} \\
    Human Baseline & \multicolumn{3}{c}{\textit{No Baseline}} & 2,574 & 85 & 3.30\% \\
    \bottomrule
    \multicolumn{7}{l}{\footnotesize $^*$Mathematical artifact of upstream API payload omissions for Claude Code in AIDev v5 (Section~\ref{sec:temporal}).}
  \end{tabular}
\end{table}

Finally, the prominent \texttt{Other} classification visible in Figure~\ref{fig:rq1_landscape}(a) (28.09\% in AIDev v5 and 20.54\% in MOSAIC-3M) reveals a disproportionate allocation of agent activity to non-functional repository maintenance. Agentic PRs modifying exclusively documentation, test files, or lockfiles---without touching application logic or infrastructure manifests---represent pure chore PRs, capturing 25.41\% (5,553) of merged agentic PRs in AIDev v5 and 17.70\% (455) in MOSAIC-3M (with the residual 2.7--2.8\% of agent PRs comprising domain-specific schemas such as Protocol Buffers and GraphQL). Across both cohorts, \texttt{README.md} is the single most frequently modified file, present in 44.70\% (2,482) of chore PRs in AIDev v5 and 25.02\% (255) in MOSAIC-3M. Consequently, non-functional maintenance chores are four to six times more prevalent than dual-scope PRs across both cohorts.

\begin{tcolorbox}[size=small,boxrule=0.8pt,sharp corners,colback=gray!5,colframe=black!80]
\textbf{Summary for RQ1:} Dual-scope PRs modifying both application code and infrastructure files account for 4.48\% (AIDev v5) and 4.55\% (MOSAIC-3M) of merged agent activity. These modifications are highly targeted, with 47.0\% to 63.9\% of instances updating a single infrastructure manifest, overwhelmingly CI/CD pipelines. Furthermore, 17.70\% to 25.41\% of agentic PRs perform pure repository maintenance chores, bypassing functional development entirely.
\end{tcolorbox}

\subsection{RQ2: Integration Latency and Contextual Bottlenecks}\label{sec:rq2}

To evaluate the integration delay induced by co-modifying infrastructure alongside application code, we apply our continuous velocity metric (merge latency) across our matched cohorts (687 AIDev v5 pairs and 80 MOSAIC-3M pairs). Our exact 1:1 matching algorithm successfully eliminated application size confounding, achieving perfect median application file equivalence between treatment and control groups (6.0 versus 6.0 files in AIDev v5; 4.0 versus 4.0 in MOSAIC-3M). Furthermore, treatment dose audits confirm that agentic infrastructure co-modifications are discrete: the median infrastructure file count within the matched treatment group is exactly 1.0 across both datasets, with 54.44\% (AIDev v5) and 61.25\% (MOSAIC-3M) of dual-scope PRs modifying a single infrastructure manifest, and active authoring confirmed in 96.51\% and 100.00\% of matched treatment units, respectively.

\begin{figure}
\footnotesize
  \includegraphics[width=0.75\linewidth]{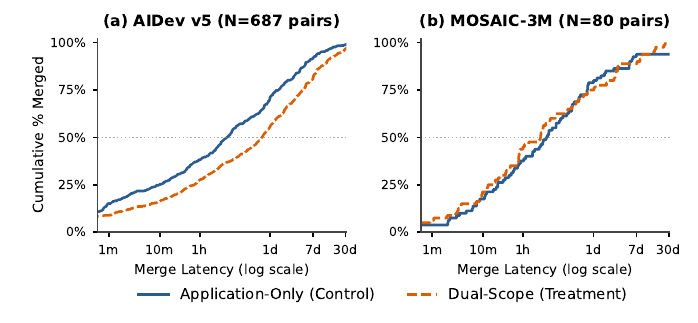}
    \caption{Empirical Cumulative Merge Latency Curves: Comparing single-domain application PRs (Control) against dual-scope PRs (Treatment) across AIDev v5 ($N = 687$ pairs) and MOSAIC-3M ($N = 80$ pairs).}
  \label{fig:rq2_survival_curves}
  \Description{Empirical cumulative distribution function (ECDF) curves comparing time-to-merge. In AIDev v5, the dual-scope curve shows a distinct horizontal shift to the right, reflecting the 13.13-hour median latency penalty. In MOSAIC-3M, the dual-scope curve rapidly reaches 50 percent within the first 31 minutes due to unreviewed silent merges, crossing the control curve.}
\end{figure}

Evaluating the macro Average Treatment Effect on the Treated (ATT), the introduction of infrastructure code increases median merge latency by +13.13 hours in AIDev v5. In this cohort, application-only control PRs merge in a median of 3.47 hours, whereas paired dual-scope treatment PRs require 16.60 hours. This latency penalty is statistically significant (Wilcoxon signed-rank $W = 80323.0$, $p = 3.55 \times 10^{-13}$, Cliff's $\delta = +0.202$, median paired difference $+0.84$ hours). Decomposing this penalty reveals that latency is strictly tied to functional integration: when agents modify core application logic alongside infrastructure (strict dual-scope), median latency reaches 19.64 hours. Conversely, when agents modify only application configurations alongside infrastructure (weak dual-scope), median latency drops to 1.13 hours, merging faster than the 3.47-hour application-only baseline. In MOSAIC-3M, the paired macro difference is not statistically significant (treatment median 2.21 hours vs. control median 2.93 hours; Wilcoxon $W = 1568.0$, $p = 0.803$, Cliff's $\delta = -0.041$). As illustrated by the empirical cumulative distribution curves in Figure~\ref{fig:rq2_survival_curves}, this cross-dataset divergence is driven by integration bypass: while AIDev v5 exhibits a persistent rightward latency shift (+13.13-hour gap at the 50th percentile), MOSAIC-3M's dual-scope curve shoots vertically to 50.00\% within the first 31 minutes (median 0.51 hours) due to silent merges without human review.

To isolate this integration penalty from baseline project scale and administrative bureaucracy, we estimate a Log-Normal parametric survival regression (AFT model) on time-to-merge, clustering Huber-White robust sandwich standard errors on matched pairs. Controlling for application file volume, code line churn, test inclusion, and repository popularity, Table~\ref{tab:rq2_bottlenecks} presents both the global treatment effect across datasets (Model 1) and the subdomain decomposition (Model 2). In Model 1, dual-scope modifications multiply total merge latency by a Time Ratio (TR) of 3.237 ($p < 0.001$) in AIDev v5. Symmetrically, reporting Model 1 for MOSAIC-3M yields a statistically non-significant treatment effect ($\text{TR} = 0.794$, $p = 0.544$), reflecting both the limited statistical power ($N=80$ pairs) and the review-bypass dynamic visualized in Figure~\ref{fig:rq2_survival_curves}.

Beyond the global treatment effect, integration latency is dictated by contextual and architectural bottlenecks. Evaluating the complete unmatched cohort at the repository level, single-domain application PRs targeting the \texttt{src/} directory merge rapidly (median 0.25 hours); however, introducing infrastructure modifications tied to this exact directory inflates median latency to 40.99 hours, representing a 163.6-fold multiplier. Similar operational friction appears in modular directories, with \texttt{apps/} and \texttt{packages/} escalating from 1.76- and 2.76-hour baselines to 66.52 hours (37.7-fold increase) and 54.01 hours (19.6-fold increase). To pinpoint the exact mechanical drivers of this friction, Model 2 in Table~\ref{tab:rq2_bottlenecks} decomposes the treatment effect via additive infrastructure subdomain indicators. Modifying Continuous Integration (CI/CD) manifests and declarative System Provisioning configurations drives the delay, multiplying merge time by Time Ratios of 3.827 ($p < 0.001$) and 2.727 ($p = 0.0070$), respectively. In contrast, procedural automation scripts and containerization configurations exhibit no statistically significant latency penalty.

\begin{table}
  \caption{Log-Normal Parametric Survival Models on PR Merge Latency}
  \label{tab:rq2_bottlenecks}
  \Description{Log-Normal AFT survival models estimating the time-to-merge multiplier for dual-scope PRs. Model 1 compares the global treatment effect across AIDev v5 and MOSAIC-3M; Model 2 decomposes the treatment effect into four infrastructure subdomains in AIDev v5, demonstrating that CI/CD manifests and system provisioning drive statistically significant integration delays.}
  \footnotesize
  \begin{tabular}{p{4.8cm} r c r}
    \toprule
    \textbf{Model / Factor} & \textbf{PR Count} & \textbf{Time Ratio [95\% CI]} & \textbf{$p$-value} \\
    \midrule
    \multicolumn{4}{l}{\textit{\textbf{Model 1: Global Treatment Effect}}} \\
    AIDev v5 Dual-Scope & 687 & 3.237 [2.545, 4.116] & $< 0.001$ \\
    MOSAIC-3M Dual-Scope & 80 & 0.794 [0.377, 1.672] & $0.544$ \\
    \midrule
    \multicolumn{4}{l}{\textit{\textbf{Model 2: Subdomain Decomposition}}} \\
    CI/CD Manifests & 532 & 3.827 [2.918, 5.020] & $< 0.001$ \\
    System Provisioning & 61 & 2.727 [1.315, 5.655] & $0.0070$ \\
    Automation Scripts & 109 & 1.522 [0.831, 2.787] & $0.1739$ \\
    Containerization & 108 & 1.132 [0.655, 1.956] & $0.6571$ \\
    \bottomrule
\multicolumn{4}{l}{\footnotesize \textit{Note:} Models control for application volume, line churn, test inclusion, and repository popularity.} \\
        \multicolumn{4}{l}{\footnotesize SEs are pair-clustered. Model 2 restricts to AIDev v5 due to sparse MOSAIC subdomains ($N \le 12$).}
  \end{tabular}
\end{table}

\begin{tcolorbox}[size=small,boxrule=0.8pt,sharp corners,colback=gray!5,colframe=black!80]
\textbf{Summary for RQ2:} Integration latency for agentic dual-scope modifications exhibits a cross-dataset divergence. In AIDev v5, infrastructure co-modifications multiply time-to-merge by a TR of 3.24 ($p < 0.001$), escalating median latency by +13.13 hours due to CI/CD ($\text{TR} = 3.83$) and system provisioning ($\text{TR} = 2.73$) bottlenecks. Conversely, in MOSAIC-3M, no significant latency penalty is observed ($\text{TR} = 0.79$, $p = 0.544$), reflecting both limited statistical power ($N=80$ pairs) and the 56.2\% of dual-scope PRs that bypass human inspection to merge silently.
\end{tcolorbox}

\subsection{RQ3: Review Friction and the Intervention Bifurcation}\label{sec:rq3}

To quantify the human intervention overhead triggered by agentic dual-scope contributions, we evaluate our sanitized metric for external maintainer behavior. Establishing the macro baseline, empirical data reveals the scale at which agentic PRs bypass human gatekeeping entirely: 79.27\% of all agentic PRs in AIDev v5 and 49.01\% in MOSAIC-3M merge silently (with exactly zero external comments and zero external review iterations). Profiling this behavior across architectures reveals that Codex accounts for 90.2\% of these unreviewed integrations in AIDev v5, while Claude drives nearly a third (31.4\%) of the overall silent activity in MOSAIC-3M.

We evaluate the causal impact of infrastructure co-modifications on human gatekeeping by contrasting the matched agent cohorts against an identically matched human baseline. To ensure baseline equivalence for this cross-sectional comparison, both the Agent and Human control groups (application-only PRs) were constrained to identical functional sizes, each exhibiting a median of exactly 4.0 application files. As detailed in Table~\ref{tab:rq3_bifurcation}, human review behavior bifurcates based on the identity of the PR author. When human developers introduce infrastructure modifications alongside application code, the external review rate remains statistically flat, shifting from 72.2\% to 73.4\%. An exact McNemar test confirms that dual-scope infrastructure modifications have no significant effect on human silent integration rates (Matched $\text{OR} = 0.85$, 95\% CI [0.38, 1.89], $p = 0.839$). 

In contrast to the stable human baseline, agentic review interventions react inconsistently to infrastructure modifications. In the AIDev v5 cohort ($N = 687$ pairs), dual-scope modifications trigger a statistically significant increase in human intervention, lowering the silent integration rate from 47.2\% to 43.1\% (Matched $\text{OR} = 0.63$, 95\% CI [0.44, 0.91], $p = 0.015$). This intervention translates to a physical workload increase, adding a mean paired difference of +0.47 external review iterations and +0.60 external comments per PR. Conversely, in MOSAIC-3M, the external review rate decreases from 52.5\% to 36.2\%, with the odds of a silent integration increasing correspondingly (Matched $\text{OR} = 2.38$, 95\% CI [1.04, 5.43], $p = 0.052$), resulting in a mean reduction of -0.19 review iterations per PR. This volatility indicates that agentic dual-scope PRs do not reliably trigger the consistent baseline of scrutiny applied to human-authored code.

\begin{table}
  \caption{Cross-Sectional Interaction: The Intervention Bifurcation}
  \label{tab:rq3_bifurcation}
  \Description{Table comparing external review intervention rates and silent integration rates across Human Baseline, MOSAIC-3M Agentic, and AIDev v5 Agentic cohorts, including McNemar Odds Ratios.}
  \footnotesize
\begin{tabular}{l cc c c}
    \toprule
\textbf{PR Scope} & \textbf{Formal Review (\%)} & \textbf{Silent Merge (\%)} & \textbf{Silent OR [95\% CI]} & \textbf{$p$-value} \\
    \midrule
    \multicolumn{5}{l}{\textbf{MOSAIC-3M Human ($N=79$)}} \\
\quad Application-Only & 72.2 & 21.5 & \multicolumn{1}{c}{1.00 (Ref)} & \multicolumn{1}{c}{---} \\
    \quad Dual-Scope & 73.4 & 19.0 & 0.85 [0.38, 1.89] & 0.839 \\
    \midrule
    \multicolumn{5}{l}{\textbf{MOSAIC-3M Agents ($N=80$)}} \\
\quad Application-Only & 52.5 & 42.5 & \multicolumn{1}{c}{1.00 (Ref)} & \multicolumn{1}{c}{---} \\
    \quad Dual-Scope & 36.2 & 56.2 & 2.38 [1.04, 5.43] & 0.052 \\
    \midrule
    \multicolumn{5}{l}{\textbf{AIDev v5 Agents ($N=687$)}} \\
\quad Application-Only & 47.5 & 47.2 & \multicolumn{1}{c}{1.00 (Ref)} & \multicolumn{1}{c}{---} \\
    \quad Dual-Scope & 53.1 & 43.1 & 0.63 [0.44, 0.91] & 0.015 \\
    \bottomrule
\multicolumn{5}{l}{\footnotesize \textit{Note:} Formal Review (\%) denotes review events; residual PRs received comments without formal reviews.} \\
\multicolumn{5}{l}{\footnotesize $p$-values reflect the Exact McNemar test ($p = 0.052$ for MOSAIC Agentic despite asymptotic CI).}
  \end{tabular}
\end{table}

To investigate whether this review bypass is an artifact of repository governance---specifically, core maintainers deploying agents to fast-track their own commits---we stratify the unmatched MOSAIC-3M cohort by organizational privilege. Baseline comparisons confirm that core maintainers (repository owners, members, and recognized collaborators) receive external reviews less frequently (50.4\%) than external contributors (71.4\%). To isolate the effect of agent identity from this administrative privilege, we analyze Stratum A, a subset of 2,232 PRs authored exclusively by external contributors with zero repository write access (i.e., contributors with no formal organizational association). Within this unprivileged group, human-authored PRs receive external review 82.7\% of the time, whereas agentic PRs are reviewed only 63.6\% of the time. A logistic regression predicting the probability of external review confirms that this deficit persists independently of administrative authority. Controlling for application volume, line churn, test inclusion, and repository popularity, the agent's identity reduces the odds of human review by 63.7\% compared to human contributors ($\text{OR} = 0.363$, 95\% CI [0.293, 0.451], $p < 0.001$). 

While Section~\ref{sec:rq2} established that CI/CD manifests cause the longest temporal delays, empirical records indicate that procedural automation scripts trigger the highest volume of human discourse. Within the matched AIDev v5 dual-scope cohort, PRs modifying automation scripts average 3.11 external review iterations and 4.24 external comments, exceeding the review volume of CI/CD pipelines (2.22 iterations, 2.93 comments). This behavioral trend is mirrored in MOSAIC-3M, where automation scripts similarly lead all categories with 3.42 external review iterations compared to just 0.88 for CI/CD manifests.

Ultimately, this intervention bifurcation dictates the integration latency explored in RQ2. Agentic dual-scope PRs that bypass gatekeeping to merge silently close in a median of just 0.43 hours (AIDev) and 0.51 hours (MOSAIC). However, when human maintainers do explicitly intervene, the median latency increases to 49.80 hours and 15.59 hours, respectively.

\begin{tcolorbox}[size=small,boxrule=0.8pt,sharp corners,colback=gray!5,colframe=black!80]
\textbf{Summary for RQ3:} Human gatekeeping behavior bifurcates based on author identity. While human review rates remain stable (72.2\% to 73.4\%) when developers co-modify infrastructure, agentic dual-scope PRs trigger volatile oversight---increasing review rates from 47.5\% to 53.1\% in AIDev v5, while dropping from 52.5\% to 36.2\% in MOSAIC-3M. Governance stratification confirms this deficit is driven by author attribution rather than administrative privilege, with agent identity reducing the odds of human review by 63.7\% ($p < 0.001$) for external contributors and allowing 43.1\% to 56.2\% of dual-scope PRs to merge silently in under 31 minutes.
\end{tcolorbox}
\section{Discussion and Implications}\label{sec:discussion}

Our empirical findings across 26,999 PRs demonstrate that while AI agents are increasingly deployed to modify operational infrastructure, modern software engineering ecosystems are structurally unequipped to safely and efficiently integrate these dual-scope contributions.

\subsection{The DevOps Integration Gap: Operational Context vs. Sandbox Benchmarks}\label{sec:discussion_devops}

Current industry benchmarks predominantly evaluate AI agents within isolated sandboxes. Whether assessing application logic (e.g., SWE-bench~\cite{swebench2024}) or specialized infrastructure automation (e.g., DevOps-Gym~\cite{tang2026devopsgym}, IaC-Eval~\cite{iaceval2024}, DevOps-Bench~\cite{devopsbench2026}), success is ultimately measured by whether the generated code passes synchronous, local validation or ephemeral environment checks. This evaluation paradigm presumes that resolving isolated tasks within sandboxes directly reflects real-world software engineering productivity. However, the empirical data from our cohorts demonstrates a fundamental shift in the engineering bottleneck. While agents rapidly author application logic, integrating dual-scope PRs introduces a structural integration trap. In the AIDev v5 cohort, this manifests as an integration delay that adds a +13.13-hour median penalty and multiplies overall time-to-merge by $3.24\times$. Conversely, in the MOSAIC-3M cohort, this bottleneck manifests as a governance bypass, with 56.2\% of agentic dual-scope PRs merging without human inspection (detailed in Section~\ref{sec:discussion_supply_chain}). This divergence indicates a productivity illusion: current agent architectures do not eliminate engineering friction; rather, they displace it from the local development environment to the repository integration queue, resulting in either stalled integrations or completely uninspected merges.

This integration friction is architecturally driven by how agents interact with specific deployment configurations. The bottleneck is most acute when functional application logic intersects with infrastructure constraints: strict dual-scope modifications---altering core source code alongside infrastructure---delay integration to a median of 19.64 hours, whereas weak dual-scope modifications targeting application configurations alone merge in just 1.13 hours. Operationally, updating a package manifest alongside a container file is a deterministic, localized syntax modification; conversely, altering core source logic while concurrently modifying CI/CD pipelines introduces complex, cross-domain state dependencies that current architectures cannot resolve without live environment feedback. Furthermore, repository metadata isolates a split in operational friction: declarative CI/CD manifests drive temporal stalling (a $3.8\times$ latency multiplier), whereas procedural automation scripts trigger the highest volume of active human review iterations (averaging 3.11 external iterations in AIDev v5 and 3.42 in MOSAIC-3M). Current architectures operate as isolated code generators, submitting functional application code directly to the repository without validating it against existing deployment workflows. While agent execution loops are designed for synchronous, local feedback, continuous integration workflows execute asynchronously on remote runners, decoupling code authoring from build verification~\cite{hilton2016usage}. Because current agents lack mechanisms to proactively ingest and resolve asynchronous remote build failures, they require human maintainers to act as the manual translation layer between the failed pipeline and the source code, inherently blocking the integration pipeline.

Consequently, this lack of deployment awareness restricts the operational scope of automated agents, driving the ``Chore Agent'' phenomenon established in Section~\ref{sec:rq1}. Between 17.70\% and 25.41\% of all agentic PRs execute pure repository maintenance, containing zero functional application logic and zero infrastructure configurations. By relegating automated agents to bumping lockfiles and updating \texttt{README.md} files (which alone comprise 25.0\% to 44.7\% of this maintenance slice), empirical evidence confirms a systemic industry bias toward isolated, low-risk execution rather than end-to-end feature deployment.

\subsection{The Intervention Bifurcation and Supply Chain Vulnerabilities}\label{sec:discussion_supply_chain}

The latency bottlenecks identified in Section~\ref{sec:rq2} are symptoms of a structural bifurcation in repository governance. As established in Section~\ref{sec:rq3}, human oversight remains stable when human developers modify infrastructure (a non-significant shift from 72.2\% to 73.4\%, $p=0.839$), but becomes highly volatile for agentic dual-scope contributions. Specifically, introducing infrastructure increases explicit review rates from 47.5\% to 53.1\% in AIDev v5, while reducing them from 52.5\% to 36.2\% in MOSAIC-3M ($p=0.052$). This bifurcation resolves the apparent cross-dataset paradox observed in Section~\ref{sec:rq2}: the latency penalty in the MOSAIC-3M cohort failed to reach statistical significance ($p = 0.544$) due to a combination of limited statistical power ($N=80$ pairs) and a structural integration bypass, wherein 56.2\% of its agentic dual-scope PRs merged silently without human review.

Without established operating procedures, repository governance defaults to a bimodal integration trap: agentic dual-scope PRs are either uninspected (merging silently in under 31 minutes) or stalled for 15.59 to 49.80 hours. This review deficit introduces a quantifiable software supply chain vulnerability. Stratification of the MOSAIC-3M cohort confirms this vulnerability persists among external contributors lacking repository write access. Within this restricted group, human-authored PRs receive explicit review 82.7\% of the time, whereas agentic PRs are reviewed 63.6\% of the time, with agent identity structurally reducing the odds of human review by 63.7\%. This disparity demonstrates that maintainer oversight is moderated by author attribution rather than patch complexity: while human-submitted PRs receive consistent inspection regardless of uninstrumented local AI assistance, explicit agent attribution induces an oversight deficit. Furthermore, permitting unreviewed modifications to continuous integration pipelines introduces a circular verification failure: because the PR alters the very workflow tasked with validating it, automated green status checks provide zero independent assurance of build integrity. In continuous integration security, industry threat models formally classify this vulnerability as Poisoned Pipeline Execution (PPE)~\cite{cider2021ppe}, where untrusted contributions manipulate workflow definitions to achieve arbitrary code execution inside build runners~\cite{koishybayev2022characterizing, muralee2023argus}. For external contributors lacking repository write access, this unreviewed pathway establishes a direct operational vector for software supply chain compromises~\cite{ladisa2023sok}.

\subsection{Actionable Implications for Tooling, Governance, and Evaluation}\label{sec:discussion_implications}

Resolving the productivity illusion requires the empirical software engineering community to update evaluation paradigms, agent architectures, and repository governance. First, benchmark designers must expand evaluation harnesses beyond isolated unit-test resolution. The community requires deployment-aware evaluation frameworks that bridge application-level issue benchmarks (e.g., SWE-bench~\cite{swebench2024}) with operational deployment benchmarks (e.g., DevOps-Bench~\cite{devopsbench2026}). Specifically, benchmark harnesses should mandate that agent patches concurrently pass functional unit suites, static manifest analyzers (e.g., \texttt{actionlint}, \texttt{hadolint}), and ephemeral execution testbeds (e.g., containerized runner jobs or Testcontainers) to verify that infrastructure modifications execute successfully before registering a resolution.

Second, AI agent architectures must shift operational pipeline verification left into their local generation loops. Rather than transmitting source code without deployment validation, agent execution runtimes must incorporate pre-submission verification by executing local containerized dry-runs (using tools like \texttt{act} to simulate continuous integration workflows locally) and running live container builds prior to opening a PR. Proactively resolving pipeline incompatibilities mitigates the protracted integration stalls (spanning 15.59 to 49.80 hours) driven by CI/CD modifications. Furthermore, the prevalence of repository maintenance agents (capturing 17.70\% to 25.41\% of agent PRs) highlights an architectural anti-pattern: organizations are deploying computationally expensive, non-deterministic LLMs for routine maintenance tasks already automated by deterministic tools such as Dependabot~\cite{he2023automating} and Renovate~\cite{renovate_tool}. Engineering teams should establish PR intake policies that categorically flag or reject agent contributions that contain zero functional application code, preserving agent compute and review capacity strictly for feature development.

Finally, neutralizing the supply chain vulnerabilities identified in Section~\ref{sec:discussion_supply_chain} requires repository maintainers to implement strict path-based access controls for automated contributors. In accordance with empirical principles of code ownership and software quality~\cite{bird2011dont}, maintainers should enforce path-based \texttt{CODEOWNERS} configurations coupled with repository rulesets that categorically disable auto-merge capabilities whenever a PR intersects with sensitive infrastructure directories (e.g., \texttt{.github/workflows/**}, \texttt{terraform/**}, \texttt{k8s/**}). Enforcing a strict branch protection policy requiring a minimum of one explicit, non-author human review sign-off on these paths neutralizes the risk of unverified build environment modifications without impeding the integration velocity of standard, application-only feature development.
\section{Threats to Validity}\label{sec:threats}

We examine the threats to the validity of our study across construct validity (measurement and operationalization), internal validity (causal identification and confounding), and external validity (generalizability and framework portability)~\cite{wohlin2012experimentation}.
% \subsection{Construct Validity}\label{sec:threats_construct}

\vspace{1mm}\noindent\textbf{Construct Validity.} Conditioning our evaluation on successfully merged PRs introduces survivorship bias, as abandoned PRs remain unobserved; consequently, our latency metrics represent a conservative lower bound of integration delay. Furthermore, because continuous integration platforms report status via GitHub's Check Runs API rather than comment threads, we evaluated total elapsed duration to capture end-to-end integration time. To accurately measure human review friction, we used a regex engine to purge automated bot notifications and explicitly filtered out author self-comments and self-reviews, which would otherwise artificially inflate intervention rates. While regex sanitation carries minor classification noise for long-tail bot accounts, it effectively isolates external maintainer behavior. Finally, governance stratification was bounded to MOSAIC-3M because AIDev v5 lacks explicit author permission metadata.

Upstream GitHub API payload limits caused missing file arrays for exceptionally large PRs in AIDev v5. We mitigated this extraction bias through cross-dataset validation against MOSAIC-3M, which maintained 99.88\% payload completeness. Furthermore, because PRs are frequently squashed upon merge (obscuring intermediate revision histories), we addressed this construct threat by measuring continuous merge latency rather than commit iterations. We calculated durations from authoritative GitHub server-side UTC timestamps to eliminate local clock drift. Additionally, our 31-day observation buffer mitigates right-censoring bias, and we floored instantaneous automated merges at 0.001 hours to prevent numerical divergence ($\log(0) = -\infty$) in our Log-Normal models.

% \subsection{Internal Validity}\label{sec:threats_internal}
\vspace{1mm}\noindent\textbf{Internal Validity.} An internal threat arises from selection bias on PR size, as dual-scope modifications inherently touch more files than single-domain tasks. We mitigated this confounding via exact multi-dimensional strata matching and caliper constraints on application file volume (detailed in Section~\ref{sec:methodology}). We matched on unique application file counts rather than raw lines of code because package lockfiles frequently introduce high-volume automated churn; line churn was instead treated as an adjusted regression covariate. To maximize common support, we sorted treatment units descending by volume prior to pairing, accepting minor variance from order-dependence. While repository-level primary language mapping in polyglot codebases risks ecosystem misclassification, concurrent matching on architectural archetypes ensures paired PRs target equivalent structural subsystems.

Metadata and baseline unobservability introduce additional confounding threats. Due to missing or unstandardized task labels across upstream datasets, we avoided subjective natural language classification. We relied on architectural archetypes and file volume as deterministic proxies, accepting latent developer intent (e.g., bug fix versus feature addition) as an unobserved confounder. Furthermore, Git records capture only terminal branch states, obscuring local prompt iterations and pre-commit refactoring. We mitigated this through exact within-agent matching, though intra-model prompting variance remains unobserved. Similarly, the MOSAIC-3M human baseline may include uninstrumented AI assistance (e.g., untracked web-based LLMs); however, this potential contamination renders our observed review deficits a conservative lower bound. To ensure statistical conclusion validity, we corrected for the paired dependence structure induced by 1:1 matching by applying cluster-robust standard errors and Pratt's zero-method. Finally, we acknowledge that observational evaluations cannot eliminate the influence of unobserved offline communication (e.g., private messaging or video calls) on integration metrics.
% \subsection{External Validity}\label{sec:threats_external}

\vspace{1mm}\noindent\textbf{External Validity.} External validity is bounded by our common support constraints and repository quality filters. Because strict caliper matching pruned PRs with outlier file volumes, our estimated ATT applies to typical PRs and does not generalize to extreme, repository-wide architectural refactorings. Furthermore, our >100 stars threshold isolates mature, collaborative open-source environments. Our findings may not generalize to personal hobby projects nor to proprietary enterprise monorepos employing dedicated Site Reliability Engineering teams, custom distributed build systems, and automated canary deployment gates.

Temporally, our June--July 2025 cohort captures current agents under controlled model versions. Rapid model evolution implies that future event-driven agent runtimes---capable of autonomously ingesting asynchronous continuous integration failures---may exhibit different integration dynamics. Additionally, evaluation tool coverage reflects inherent dataset asymmetries (e.g., Cursor is exclusive to AIDev v5, while the human baseline is exclusive to MOSAIC-3M). Finally, while our deterministic taxonomy and matching framework provide a reusable methodology, applying them requires tuning directory path assumptions for non-standard monorepos, securing sufficiently large cohorts to survive the curse of dimensionality, and monitoring upstream API payload limits on large PRs.
\section{Related Work}\label{sec:related_work}

We situate our findings across three domains of software engineering research. Section~\ref{sec:related_benchmarks} contrasts isolated agentic benchmarks with real-world deployment environments. Section~\ref{sec:related_mining} compares our causal matching methodology against prior empirical repository mining studies. Section~\ref{sec:related_governance} connects our observations to modern code review, automated reliance, and continuous integration security.

\subsection{From Execution Sandboxes to Socio-Technical Deployment Context}\label{sec:related_benchmarks}

Evaluation frameworks for coding agents have shifted from isolated function completion~\cite{chen2021evaluating} to repository-scale issue resolution (e.g., SWE-bench~\cite{swebench2024}, DevBench~\cite{li2024devbench}). In parallel, specialized operations benchmarks like DevOps-Gym~\cite{tang2026devopsgym} and IaC-Eval~\cite{iaceval2024} evaluate infrastructure tasks in isolation. Despite differing scopes, these frameworks share a fundamental design: they test agents inside closed sandboxes against pre-packaged, synchronous unit tests or static linters.

While sandboxes ensure reproducibility, they create a blind spot by separating code generation from asynchronous integration pipelines. Real-world software engineering requires the continuous co-evolution of application logic and deployment infrastructure~\cite{jiang2015coevolution}. By evaluating application and infrastructure code in disjoint tracks, current benchmarks treat deployment configurations as separate coding exercises rather than the mandatory operational context of a feature. This mirrors the test-suite overfitting problem in automated program repair~\cite{smith2015cure}, where patches passing local tests fail to generalize beyond the immediate test oracle.

Our study bridges this gap by evaluating agents in live repositories. We reveal that sandbox success can mask a productivity illusion: while agents generate code quickly, integrating dual-scope contributions introduces friction. In environments where continuous integration executes asynchronously, co-modifying infrastructure multiplies time-to-merge by $3.24\times$. This delay is driven heavily by continuous integration manifests (a $3.8\times$ penalty), which require remote validation that local sandboxes miss. By isolating this integration cost, we show that agent evaluation must expand beyond local pass rates to account for real-world deployment dependencies.

\subsection{Empirical Mining of Automated Code: Rule-Based Bots to Causal Agent Analytics}\label{sec:related_mining}

Empirical PR mining relies on foundational integration metrics such as merge latency and code churn~\cite{gousios2014exploratory}. Historically, automated bots (e.g., Repairnator~\cite{urli2018repairnator} and review tools~\cite{wessel2020effects}) introduced maintainer triage hurdles and communication noise. Despite industry narratives framing LLM agents as autonomous junior developers, our empirical evidence reveals strong parallels to these legacy tools. As shown in RQ1, agents frequently operate as ``Chore Agents'': 17.70\% to 25.41\% of merged agentic PRs execute pure non-functional maintenance rather than synthesizing features.

Recent releases of large-scale agent corpora~\cite{aidev_v5,mosaic3m_2026} have spurred empirical analyses of modern agents. Yoshioka et al.~\cite{yoshioka2026} and Peralta et al.~\cite{peralta2026} analyzed raw agent merge outcomes and reviewer workflows, while Ehsani et al.~\cite{ehsani2026} observed that agent PRs modifying continuous integration pipelines face elevated failure rates. However, analyzing unadjusted populations conflates integration complexity with PR size. Because dual-scope PRs inherently modify more files, naive observational comparisons cannot distinguish whether delays stem from infrastructure inclusion or simply larger code volume.

Our study introduces causal identification to isolate this marginal integration cost. We discard subjective natural language task labels---which collapsed in coverage from 66.32\% in July 2025 to just 1.13\% by August 2025 in AIDev v5---in favor of a deterministic four-tier file taxonomy. To eliminate size-based selection bias, we apply 1:1 nearest-neighbor matching under conditional exchangeability~\cite{austin2011psm}. By pairing PRs within exact four-dimensional strata (Agent $\times$ Architectural Archetype $\times$ Test Verification $\times$ Language Ecosystem) and enforcing an $\epsilon \le 0.28$ caliper on application file volume, we transition agent analytics from unadjusted correlations to isolated, causally identified integration metrics.

\subsection{Modern Code Review, Automated Reliance, and Circular CI/CD Verification}\label{sec:related_governance}

Modern code review functions as a socio-technical process for knowledge sharing and risk mitigation rather than strictly defect detection~\cite{bacchelli2013expectations}. As repositories scale, maintainers increasingly rely on automated continuous integration checks to accelerate integration~\cite{vasilescu2015quality}. Our findings demonstrate how this automated reliance introduces a structural vulnerability when applied to agents. While human developer review rates remain stable when co-modifying infrastructure, agentic dual-scope PRs trigger an intervention bifurcation. Explicitly reviewed agentic PRs stall for 15.59 to 49.80 hours, while 43.1\% to 56.2\% bypass human inspection and are silently integrated.

Because repository integrators prioritize PRs based heavily on contributor track record and project affiliation~\cite{gousios2015integrator}, we must isolate administrative privilege to explain these silent merges. To determine whether silent integrations represent repository owners merging their own commits without peer review, we stratified contributions by governance privilege. This oversight deficit persists independently of administrative authority: even among unprivileged external contributors, agent identity independently reduces the odds of human review by 63.7\%.

Permitting unverified modifications to deployment infrastructure introduces quantifiable supply-chain vulnerabilities. Continuous integration environments are highly vulnerable to unvetted runner script executions~\cite{koishybayev2022characterizing} and untrusted input manipulations~\cite{muralee2023argus}. Industry threat models classify this vector as Poisoned Pipeline Execution (PPE)~\cite{cider2021ppe}, where attackers manipulate configurations to compromise build pipelines. Our findings identify a socio-technical mechanism that exacerbates this vulnerability: the circular verification paradox. When an agent co-modifies the continuous integration pipeline (e.g., \texttt{.github/workflows/*}) and the PR merges without formal review, the verification oracle is compromised. Because the patch altered the validation logic tasked with verifying it, the passing automated check provides zero independent assurance of build integrity. By quantifying the high rate of silent integration, our study establishes that current repository workflows provide an uninspected vector for pipeline compromises.

\section{Conclusion}\label{sec:conclusion}

While current benchmarks evaluate coding agents inside isolated sandboxes, deploying features in collaborative repositories requires modifying continuous integration pipelines, containers, and provisioning manifests alongside application logic. To isolate this integration overhead without relying on natural language task labels (which decayed from 66.32\% to 1.13\% coverage in upstream records), we evaluated 26,999 PRs across AIDev v5 and MOSAIC-3M, covering six agent architectures and an aligned human baseline. We contribute a deterministic four-tier file taxonomy and an exact 4D strata-matching design as a reusable empirical framework to evaluate cross-domain contributions under conditional exchangeability.

Our findings reveal that agents operate primarily as application coders or maintenance workers rather than DevOps engineers: pure maintenance tasks capture 17.70\% to 25.41\% of agent activity, whereas dual-scope PRs represent only 4.48\% to 4.55\% of agent contributions. When agents co-modify infrastructure, repositories default to a bimodal integration trap. When maintainers explicitly review these contributions, dual-scope modifications add a median delay of +13.13 hours and multiply time-to-merge by 3.24× in AIDev v5, driven by continuous integration manifests (3.8×) and system provisioning (2.7×). Conversely, maintainer gatekeeping exhibits a structural bypass: controlling for permissions, agent identity reduces the odds of human review by 63.7\% for external contributors, allowing 43.1\% to 56.2\% of dual-scope PRs to merge silently in under 31 minutes.

Closing this gap requires agent runtimes to execute local container dry-runs prior to submission, while maintainers must enforce path-based \texttt{CODEOWNERS} rules on workflow directories. Finally, our findings establish a three-fold empirical agenda: (1) constructing deployment-aware benchmarks that evaluate agents against ephemeral continuous integration environments rather than isolated sandboxes; (2) applying our matching framework to unmerged PRs to measure PR rejection and abandonment rates; and (3) conducting qualitative field studies to unpack why maintainers systematically bypass human review for automated dual-scope contributions.
\section*{Data Availability}\label{sec:data_availability}

To support scientific transparency and independent replication, our complete replication package is publicly accessible via Anonymous GitHub at:
\begin{center}
\url{https://anonymous.4open.science/r/agentic-PR-FSE2027-52B0}
\end{center}
The repository provides: (1) \texttt{pipeline\_utils.py}, implementing the deterministic four-tier file taxonomy classifier, automated Hugging Face dataset caching, and greedy 1:1 exact strata matching; (2) three documented Jupyter notebooks reproducing all empirical analyses, statistical regressions, and findings across RQ1--RQ3; (3) an environment manifest (\texttt{requirements.txt}); and (4) a \texttt{README.md} detailing step-by-step reproduction instructions.

Upon manuscript acceptance, the package will be de-anonymized, permanently archived on Zenodo with a persistent Digital Object Identifier (DOI), and released under open-source terms (MIT License for code and CC-BY 4.0 for data).

% %%
% %% The next two lines define the bibliography style to be used, and
% %% the bibliography file.
\bibliographystyle{ACM-Reference-Format}
\bibliography{sample-base}

% %%
% %% If your work has an appendix, this is the place to put it.
% \appendix

\end{document}